\documentclass[twocolumn]{aastex61}
\date{\today}
\begin{document}

\title{Median Statistics Analysis of Deuterium Abundance Measurements and Spatial Curvature Constraints}

\author{Jarred Penton}
\affiliation{Department of Physics, Kansas State University, 116 Cardwell Hall, Manhattan, KS 66506, USA}
\affiliation{Department of Physics, Fort Hays State University, 225 Tomanek Hall, Manhattan, KS 67601, USA}

\author{Jacob Peyton}
\affiliation{Department of Physics, Kansas State University, 116 Cardwell Hall, Manhattan, KS 66506, USA}

\author{Aasim Zahoor}
\affiliation{Department of Physics, Kansas State University, 116 Cardwell Hall, Manhattan, KS 66506, USA}
\affiliation{Department of Mechanical Engineering, National Institute of Technology - Karnataka, NH 66, Srinivas Nagar, Surathkal, Mangaluru, Karnataka 575025, India}

\author{Bharat Ratra}
\affiliation{Department of Physics, Kansas State University, 116 Cardwell Hall, Manhattan, KS 66506, USA}

\begin{abstract}
\cite{Zavarygin2018} compiled a list of 15 deuterium abundance measurements, discarded two because the remaining 13 measurements are then consistent with gaussianity, and found that the weighted mean baryon density ($\Omega_b h^2$) determined from the 13 measurements is mildly discrepant ($1.6\sigma$) with that determined from the Planck 2015 cosmic microwave background anisotropy data in a flat cosmogony. We find that a median statistic central estimate of $\Omega_b h^2$ from all 15 deuterium abundance measurements is a more accurate estimate, is very consistent with $\Omega_b h^2$ estimated from Planck 2015 data in a flat cosmogony, but is about $2\sigma$ lower than that found in a closed cosmogonical model from the Planck 2015 data. 
\end{abstract}


\section{\label{intro}\textbf{Introduction}}

By the time the Universe was a few minutes old, the strong force had fused together neutrons and protons and synthesized the light nuclei. In the standard cosmological model the primordial light nuclei abundances depend only on $\Omega_b h^2$ (here $\Omega_b$ is the baryonic matter density parameter and $h$ is the Hubble constant in units of 100 km $\rm{s}^{-1}$ $\rm{Mpc}^{-1}$). Consequently, primordial light nuclei abundance measurements can be used to determine $\Omega_b$, and $\Omega_b$ determined using different light nuclei must agree, if the standard model is correct.

Deuterium is particularly valuable in this regard as its predicted abundance is quite sensitive to the value of $\Omega_b h^2$. Spectroscopic analysis of absorption of quasar light by foreground low-metallicity gas clouds is used to estimate the primordial deuterium to hydrogen abundance, $\mathrm{(D/H)}_{\rm{p}}$.\footnote{The analysis focuses on Lyman absorption lines produced by the gas clouds. The deuterium Lyman lines are at slightly shorter wavelengths than those of hydrogen. The difference in absorption at these two sections of wavelengths determines D/H.} \cite{Zavarygin2018}, hereafter Z18, have compiled a list of 15 $\mathrm{(D/H)}_{\rm{p}}$ measurements.

With the assumption of a cosmological model, $\Omega_b h^2$ can also be determined by fitting the cosmological model to observational data, such as cosmic microwave background (CMB) anisotropy measurements \citep{PlanckCollab2016b}. Comparisons between $\Omega_b h^2$'s determined from different data is a particularly compelling test of the standard cosmological model, and this can also be used to constrain other cosmological model parameters.

Z18 argue that two of their 15 measurements are outliers, if their 15 measurements were drawn from a Gaussian distribution. Discarding these two measurements Z18 determine a weighted mean $\Omega_b h^2$ using the remaining 13 deuterium abundance measurements. They note that this value differs at $1.6\sigma$ from that determined by using the Planck 2015 TT + lowP + lensing CMB anistropy data \citep{PlanckCollab2016b}.

Non-gaussian data compilations are not that rare \citep{Bailey2017}. Well-known examples include Hubble constant measurements \citep{Chen2003, Bethapudi2017, Zhang2018}, $^7 \rm{Li}$ abundance data \citep{Crandall2015a, Zhang2017}, LMC and SMC distance observations \citep{de Grijs2014, Crandall2015b}, and the Milky Way $R_0$ and $\Theta_0$ parameter measurements \citep{de Grijs2016, Camarillo2018a, de Grijs2017, Rajan2018, Camarillo2018b}. Since gaussianity is assumed in parameter estimation \citep[e.g.,][]{Samushia2007, Samushia2010, Farooq2015}, much effort has been devoted to testing for intrinsic non-gaussianity \citep[][and references therein]{Park2001, PlanckCollab2016a}, as distinct from non-gaussianity introduced by the measurement procedure.

Conventional techniques cannot be used to analyze data with non-gaussian errors \citep{Gott2001, Bailey2017}; this was one of the motivations for the development of median statistics \citep{Gott2001}. Median statistics does not make use of the errors on individual measurements and so is not affected by incorrect errors. On the other hand, since it does not use this information it is less constraining than a weighted mean analysis.

Perhaps the most well known example of  the use of median statistics is its application to the measurement of the Hubble constant \citep{Gott2001, Chen2003, Chen2011, Rajan2018}. In this paper we apply median statistics to Z18's compilation of 15 $\mathrm{(D/H)}_{\rm{p}}$ measurements.

We first examine the gaussianity of the Z18 data compilation. In agreement with Z18, we find that the full 15 measurements data set is non-gaussian, while their favored truncated set of 13 measurements is consistent with gaussianity. We then argue that the less precise median statistics summary estimate for $\Omega_b h^2$ for all 15 measurements is a more accurate representation of the data than is the more precise weighted mean summary estimate for the truncated data set of 13 measurements. We find that the median statistics $\Omega_b h^2$ determined from $\mathrm{(D/H)}_{\rm{p}}$ measurements is very consistent with those determined from other cosmological data in the context of spatially-flat cosmogonies, but is about 2$\sigma$ lower than those determined from cosmological data when using non-flat cosmogonies.

\section{\label{sec:2}\textbf{Data}}

Z18 have collected 15 D/H measurements from various sources. These are listed in Table~\ref{tab:1} of our paper. Table 5 of Z18 provides more information about these measurements. 

Z18 note that the ``scatter in D/H measurements exceeds that expected on the basis of the statistical error estimates." They analyze this compilation of 15 D/H measurements using a modified Least Trimmed Squares (LTS) procedure, chosen to discard the two most deviant of these 15 measurements. This procedure identifies the \cite{Pettini2001} and \cite{Srianand2010} measurements as the outliers. Z18 find the weighted mean of the remaining 13 measurements is $\mathrm{(D/H)}_{\rm{p}} = (2.545 \pm0.0025) \times10^{-5}$ ($1\sigma$ error; we find 2.544 instead of 2.545) and the corresponding reduced $\chi^2$ is unity.

\begin{table}[h]
\caption{D/H measurements from Z18}
\begin{tabular}{lll} \hline\hline
\multicolumn{1}{c}{Quasar}&\multicolumn{1}{c}{D/H$(\times10^5)$}&\multicolumn{1}{c}{References} \\ \hline
HS 0105+1619&$2.58^{+0.16}_{-0.15}$&\cite{Cooke2014} \\
J0407-4410&$2.8^{+0.8}_{-0.6}$&\cite{Noterdaeme2012} \\
Q0913+072&$2.53^{+0.11}_{-0.10}$&\cite{Cooke2014} \\
Q1009+2956&$2.48^{+0.41}_{-0.13}$&\cite{Zavarygin2018} \\
J1134+5742&$2.0^{+0.7}_{-0.5}$&\cite{Fumagalli2011} \\
Q1243+3047&$2.39\pm{0.08}$&\cite{Cooke2018} \\
J1337+3152&$1.2^{+0.5}_{-0.3}$&\cite{Srianand2010} \\
SDSS&$2.62\pm{0.07}$&\cite{Cooke2016} \\
J1358+6522&$2.58\pm{0.07}$&\cite{Cooke2014} \\
J1419+0829&$2.51\pm{0.05}$&\cite{Cooke2014} \\
J1444+2919&$1.97^{+0.33}_{-0.28}$&\cite{Balashev2016} \\
J1558-0031&$2.40^{+0.15}_{-0.14}$&\cite{Cooke2014} \\
PKS1937-1009&$2.45^{+0.30}_{-0.27}$&\cite{Riemer-Sorensen2015} \\
PKS1937-101&$2.62\pm{0.05}$&\cite{Riemer-Sorensen2017} \\
Q2206-199&$1.65\pm{0.35}$&\cite{Pettini2001} \\ \hline
\end{tabular}
\label{tab:1}
\end{table}

\section{\label{sec:3}\textbf{Analysis}}

Z18 note that at least one of the 15 measurements listed in Table~\ref{tab:1} has smaller error bars than it should have if the measurements were drawn from a Gaussian distribution. In this Section we quantitatively confirm that the 15 measurements (All 15) are distributed non-gaussianly, while the truncated set of 13 (Truncated 13) favored by Z18 are gaussianly distributed. We then argue that a median statistics determination of a central estimate and error bars of the All 15 set is a better estimate of $\mathrm{(D/H)}_{\rm{p}}$ than is the Z18 LTS weighted mean and error bar of the Truncated 13 set.

To study the gaussianity of a data compilation we need to use a central estimate of the data. We consider three here: the median, weighted mean, and arithmetic mean central estimates. 

Median statistics does not make use of the individual measurement errors. Consequently, the uncertainty of the summary central value (the median) determined using median statistics will be larger than that of the weighted mean (which makes use of the individual measurement errors). The median is defined as the value with half of the individual measurements above it and half below it. \cite{Gott2001} showed that for $M_i$, $i = 1, 2, ... , N$, independent measurements, the probability of the true median being placed between measurements $M_i$ and $M_{i+1}$ is 
\begin{equation}
P = \frac{2^{-N}N!}{i!(N-i)!}.
\label{eq:1}
\end{equation}
The $1\sigma$ ($2\sigma$) error range about the median is the range that includes 68.27\% (95.45\%) of the probability under $P$. For asymmetric distributions, upper and lower error bars are computed.

The weighted mean has the benefit of using the individual measurement errors with the risk that some of them might be inaccurate \citep{Podariu2001}. For $M_i \pm \sigma_i$ measurements with errors $\sigma_i$, the weighted mean central value is
\begin{equation}
M_{\rm{wm}} = \frac{\sum_{i=1}^NM_i/\sigma^2_i}{\sum^N_{i=1}1/\sigma^2_i},
\label{eq:2}
\end{equation}
\noindent and the weighted mean standard deviation is
\begin{equation}
\sigma_{\rm{wm}} = \frac{1}{\sqrt{\sum^N_{i=1}1/\sigma^2_i}}.
\label{eq:3}
\end{equation}
\noindent We also use the arithmetic mean central estimate
\begin{equation}
M_{\rm{m}} = \frac{1}{N}\sum_{i=1}^NM_i,
\label{eq:4}
\end{equation}
\noindent with standard deviation
\begin{equation}
\sigma_{\rm{m}} = \sqrt{\frac{1}{N^2}\sum^N_{i=1}(M_i-M_{\rm{m}})^2}.
\label{eq:5}
\end{equation}

Table~\ref{tab:2} lists these central estimates with $1\sigma$ error bars for the D/H data in Table~\ref{tab:1}. The weighted mean error is half as big as the smallest individual error in Table~\ref{tab:1} while the symmetrized median error is about 30\% larger than the smallest individual error in Table~\ref{tab:1}.

\begin{table}[h]
\caption{Central estimates and $1\sigma$ error bars for D/H ($\times10^5$) measurements}
\centering
\begin{tabular}{lll} \hline\hline
Central Estimate&Truncated 13 & All 15 \\ \hline 
Median & $2.51^{+0.07}_{-0.06}$ & $2.48^{+0.05}_{-0.08}$ \\ 
Arithmetic mean & $2.456\pm0.063$ & $2.32\pm0.108$ \\ 
Weighted mean & $2.544\pm0.025$ & $2.53\pm0.025$ \\ \hline
\end{tabular}

\label{tab:2}
\end{table}

\subsection{\label{sec:3.1}Error Distributions} 

The central estimates are used to construct error distributions. For each central estimate and uncertainty, $M_{\rm{CE}} \pm \sigma_{\rm{CE}}$, a new error distribution data set is created by utilizing 
\begin{equation}
N_{\sigma_i} = \frac{M_i - M_{\rm{CE}}}{\sqrt{\sigma^2_i + \sigma^2_{\rm{CE}}}}.
\label{eq:6}
\end{equation}  
\noindent This formula assumes that the central estimate is not correlated with the data.

For gaussianly distributed measurements and for a weighted mean central estimate computed from these measurements, the pull that correctly accounts for the correlations is
\begin{equation}
N_{\rm{wm^{-}}} = \frac{M_i - M_{\rm{wm}}}{\sqrt{\sigma^2_i - \sigma^2_{\rm{wm}}}}.
\label{eq:7}
\end{equation}
\noindent See the Appendix of \cite{Camarillo2018a} for a derivation of this expression.

To simplify the analysis, this new error distribution data set is then symmetrized about 0.\footnote{This is done by copying its negative back into itself and dividing by 2, creating a symmetric distribution centered at 0, with standard deviation equal to 1.} An error distribution is created for each central estimate; these are used for gaussianity tests.

\subsection{\label{sec:3.2}Gaussianity Tests} 

\begin{table}
\caption{KS Test Probabilities}
\begin{tabular}{ccccccc} \hline\hline
&\multicolumn{2}{c}{Truncated 13}&&\multicolumn{2}{c}{All 15}& \\ \hline	
Dist.&$S$\tablenotemark{a}&$p$\tablenotemark{b}&$n$\tablenotemark{c}&$S$\tablenotemark{a}&$p$\tablenotemark{b}&$n$\tablenotemark{c} \\ \hline
\multicolumn{7}{l}{\textbf{Median}} \\ 
\multicolumn{1}{l}{Gaussian}&1&$0.999$&&1&$0.809$& \\
&$0.867$&$0.999$&&$1.269$&$0.999$& \\
\multicolumn{1}{l}{Laplacian}&1&$0.943$&&1&$0.869$& \\
&$0.796$&$0.999$&&$1.306$&$0.996$& \\
\multicolumn{1}{l}{Cauchy}&1&$0.385$&&1&$0.921$& \\
&$0.464$&$0.991$&&$0.836$&$0.981$& \\
\multicolumn{1}{l}{Student's $t$}&1&$0.999$&$2000$&1&$0.992$&$2$ \\
&$0.76$&$0.999$&$3$&$1.25$&$0.999$&$28$ \\
\multicolumn{7}{l}{\textbf{Weighted Mean +}} \\
\multicolumn{1}{l}{Gaussian}&1&$0.999$&&1&$0.885$& \\
&$1.012$&$0.999$&&$1.224$&$0.999$& \\
\multicolumn{1}{l}{Laplacian}&1&$0.997$&&1&$0.999$& \\
&$0.926$&$0.999$&&$1.162$&$0.999$& \\
\multicolumn{1}{l}{Cauchy}&1&$0.517$&&1&$0.948$& \\
&$0.496$&$0.992$&&$0.672$&$0.999$& \\
\multicolumn{1}{l}{Student's $t$}&1&$0.999$&$22$&1&$0.999$&$2$ \\
&$1.01$&$0.999$&$111$&$0.94$&$0.999$&$2$ \\
\multicolumn{7}{l}{\textbf{Weighted Mean $-$}} \\
\multicolumn{1}{l}{Gaussian}&1&$0.997$&&1&$0.613$& \\
&$1.093$&$0.999$&&$1.464$&$0.999$& \\
\multicolumn{1}{l}{Laplacian}&1&$0.999$&&1&$0.966$& \\
&$0.995$&$0.999$&&$1.316$&$0.999$& \\
\multicolumn{1}{l}{Cauchy}&1&$0.604$&&1&$0.950$& \\
&$0.531$&$0.993$&&$0.682$&$0.999$& \\
\multicolumn{1}{l}{Student's $t$}&1&$0.999$&$7$&1&$0.999$&$2$ \\
&$0.96$&$0.999$&$4$&$1.11$&$0.999$&$2$ \\
\multicolumn{7}{l}{\textbf{Arithmetic Mean}} \\ 
\multicolumn{1}{l}{Gaussian}&1&$0.999$&&1&$0.238$& \\
&$1.005$&$0.998$&&$1.949$&$0.981$& \\
\multicolumn{1}{l}{Laplacian}&1&$0.994$&&1&$0.338$& \\
&$0.991$&$0.995$&&$1.731$&$0.876$& \\
\multicolumn{1}{l}{Cauchy}&1&$0.612$&&1&$0.722$& \\
&$0.629$&$0.965$&&$1.05$&$0.770$& \\
\multicolumn{1}{l}{Student's $t$}&1&$0.999$&$69$&1&$0.722$&$1$ \\
&$1.00$&$0.998$&$69$&$1.94$&$0.981$&$123$ \\ \hline
\end{tabular}
\tablenotetext{a}{The scale factor $S$ fixed to 1 or that which maximizes $p$.}
\tablenotetext{b}{The probability ($p$-value) that the data doesn't not come from the PDF.}
\tablenotetext{c}{Student's $t$ distribution parameter $n$.}
\label{tab:3}
\end{table} 

\begin{table*}
\caption{Comparison between $\Omega_b h^2$ determined from (D/H)$_{\rm{p}}$ and from CMB anisotropy and other data}
\centering
\begin{tabular}{ccccccc} \hline\hline
&\multicolumn{3}{c}{CMB data alone}&\multicolumn{3}{c}{CMB and other data} \\
Cosmogony&$\Omega_b h^2$&WM $\sigma$&Median $\sigma$&$\Omega_b h^2$& WM $\sigma$&Median $\sigma$ \\ \hline			
Flat $\Lambda$CDM&0.02225 $\pm0.00023$&1.5&0.34&0.02232 $\pm0.00019$&1.8&0.51 \\
Nonflat $\Lambda$CDM&0.02305 $\pm0.0002$&4.1&2.1&0.02305 $\pm0.00019$&4.1&2.1 \\
Flat XCDM&0.02229 $\pm0.00023$&1.6&0.43&0.02233 $\pm0.00021$&1.8&0.52 \\
Nonflat XCDM&0.02305 $\pm0.0002$&4.1&2.1&0.02305 $\pm0.0002$&4.1&2.1 \\
Flat $\phi$CDM&0.02221 $\pm0.00023$&1.4&0.26&0.02238 $\pm0.0002$&2.0&0.64 \\
Nonflat $\phi$CDM&0.02303 $\pm0.0002$&4.0&2.1&0.02304 $\pm0.0002$&4.0&2.1 \\ \hline
\end{tabular}

\label{tab:4}
\end{table*}

In this subsection we study the two data sub-compilations of Table~\ref{tab:1}, the All 15 set and the Truncated 13 set which excludes the \cite{Pettini2001} and \cite{Srianand2010} measurements, to determine whether they are non-gaussian or gaussian. We do this by comparing their error distributions to four widely used distributions. 

The first distribution we consider is the Gaussian distribution, defined with a mean of zero, and standard deviation of 1. Its probability density function (PDF) is
\begin{equation}
P(|\textbf{N}|)=\frac{1}{\sqrt{2\pi}}\exp(-|\textbf{N}|^{2}/2),
\label{eq:8}
\end{equation}
\noindent with $68.27\%$ ($95.45\%$), or $1\sigma$ ($2\sigma$), of the probability lying within $|N| \leq 1$ $( |N| \leq 2)$.

The second distribution we compare the error distributions to is the Laplace, or Double Exponential, distribution. It is characterized by a sharp peak, and longer tails than a Gaussian distribution
\begin{equation}
P(|\textbf{N}|)=\frac{1}{2}\exp{(-|\textbf{N}|)},
\label{eq:9}
\end{equation}
\noindent with $68.27\%$ $(95.45\%)$, or $1\sigma$ ($2\sigma$), of the probability lying within $|N| \leq 1.2$ $( |N| \leq 3.1)$.

The third distribution we use is the Cauchy, or Lorentz, distribution. It's shaped similarly to the Gaussian distribution, but has longer and thicker tails, with $68.27\%$ $(95.45\%)$, or $1\sigma$ ($2\sigma$), of the probability lying within $|N| \leq 1.8$ $( |N| \leq 14)$. It is described as
\begin{equation}
P(|\textbf{N}|)=\frac{1}{\pi}\frac{1}{1+|\textbf{N}|^{2}}.
\label{eq:10}
\end{equation}

The final distribution is the Student's $t$ distribution. It's centered around 0, and has an additional parameter, a positive integer $n$. At $n = 1$ this distribution is the Cauchy distribution, and as $n$ approaches infinity it approaches the Gaussian distribution. The Student's $t$ PDF is
\begin{equation}
P(|\textbf{N}|)=\frac{\Gamma[(n+1)/2]}{\sqrt{\pi n}\Gamma(n/2)}\frac{1}{(1+|\textbf{N}|^{2}/n)^{(n+1)/2}}.
\label{eq:11}
\end{equation}

In addition to the standard forms of the PDFs in Eqs.~(\ref{eq:8}) --- (\ref{eq:11}), we also consider scaled distributions where we replace $|\bf{N}|$ in the formulae above by $|\bf{N}|$$/S$ where $S$ is the scale factor.

In our analyses here we allow $S$ to vary in steps of $0.001$ from 0.001 to $2.5$. For the Student's $t$ distribution we lower the number of steps by varying $S$ in steps of $0.01$, and also vary $n$ from 1 to 2000 in steps of 1.

We use the Kolmogorov-Smirnov (KS) test to compare the error distributions to the PDFs. The KS test utilizes the $D$-statistic, which is the largest difference between the cumulative distribution function of the error distribution and of the PDF under consideration. The $D$-statistic is then used in the inverted Kolmogorov distribution, to determine the $p$-value
\begin{equation}
p = 2\sum_{i=1}^{\infty}({-1})^{i-1}e^{-i^2z^2},
\label{eq:12}
\end{equation} 
\noindent where
\begin{equation}
z = \Big(\sqrt{N} + 0.12 + \frac{0.11}{\sqrt{N}}\Big)D.
\label{eq:13}
\end{equation}
\noindent The $p$-value is the probability that the $D$-statistic could be smaller than measured given a similar data set. In general terms, the higher the $p$-value the more similar the two distributions. More precisely, the $p$-value is the probability that the data doesn't not come from the PDF it is being compared to. Our PDF comparison results are listed in Table~\ref{tab:3}.

Focusing first on the All 15 data set, we see that the Gaussian distribution is not a reasonable fit unless the scale factor is pulled away from unity. This agrees with the Z18 finding that at least one D/H measurement in Table~\ref{tab:1} has smaller error bars than expected if they were drawn from a Gaussian distribution. The other All 15 results in Table~\ref{tab:3} confirm that the All 15 data set is non-gaussian. Considering the Truncated 13 data results in Table~\ref{tab:3}, we see that they are quite consistent with gaussianity, as found by Z18. This is a consequence of the removal of the outlying \cite{Pettini2001} and \cite{Srianand2010} measurements. We emphasize however, that there is no guarantee that these two measurements are incorrect. All we know for sure is that in the All 15 compilation the error bars are such that these 15 measurements cannot have been drawn from a Gaussian distribution. In such a situation it is best to use a median statistics estimate of the summary central value and error bars, instead of using an approach such as LTS that results in a more precise summary central estimate that might possibly be less accurate.

\subsection{\label{sec:3.3}Baryonic Density Measurements and Spatial Curvature}

By using the \cite{Coc2015} equation given in Z18 we have
\begin{equation}
\Omega_b h^2 = 0.02225\Bigg[\frac{(2.45 \pm 0.04) \times 10^{-5}}{(D/H)_{\rm{p}}}\Bigg]^{1/1.657}.
\label{eq:15}
\end{equation}
To the Lyman absorption error bar on $\Omega_b h^2$ derived using this equation, we add an additional $\pm0.00021$ nuclear data uncertainty \citep{Coc2015} in quadrature. 

Using the All 15 median central estimate of $(D/H)_{\rm{p}} = (2.48 \pm 0.065)\times10^{-5}$, with symmetrized error, we get $\Omega_b h^2 = 0.02209 \pm 0.00041$. The weighted mean $(D/H)_{\rm{p}} = (2.544 \pm 0.025)\times10^{-5}$ for the Truncated 13 data set results in $\Omega_b h^2 = 0.02175 \pm 0.00025$, in good agreement with Z18's $\Omega_b h^2 = 0.02174 \pm0.00025$. The median error bar on $\Omega_b h^2$ is 65\% larger than that on the weighted mean and the median and weighted mean $\Omega_b h^2$ central estimates differ by $0.7\sigma$ (of the quadrature sum of their errors).

CMB anisotropy and other cosmological data can be used to determine $\Omega_b h^2$, given a cosmogonical model. $\Omega_b h^2$ determined using such data is sensitive to the geometry of space, being larger in the nonflat, closed models \citep{Ooba2018a, Ooba2017a, Ooba2017b, Ooba2018b, Park2018a, Park2018b, Park2018c}. The second and fifth columns of Table~\ref{tab:4} list $\Omega_b h^2$ values determined assuming the cosmogonies listed in the first column of the table \citep{Park2018b, Park2018c}. The second column lists values determined using Planck 2015 TT + lowP + lensing CMB anisotropy data \citep{PlanckCollab2016b}, while the fifth column lists $\Omega_b h^2$ determined using this CMB data in conjunction with the latest Type Ia supernova apparent magnitude measurements, baryon acoustic oscillation observations, Hubble parameter data, and growth factor measurements \citep{Park2018b, Park2018c}. 

The third, fourth, sixth, and seventh columns of Table~\ref{tab:4} list the number of standard deviations between the cosmology-determined $\Omega_b h^2$ listed in the second and fifth columns and the (D/H)$_{\rm{p}}$-determined $\Omega_b h^2$, for the Truncated 13 weighted mean analysis (third and sixth columns) and for the All 15 median analysis (fourth and seventh columns), in multiples of $\sigma$ determined by adding in quadrature the cosmology and (D/H)$_{\rm{p}}$ uncertainties on the two $\Omega_b h^2$'s. 

Table~\ref{tab:4} shows that for the flat models the Truncated 13 weighted mean (D/H)$_{\rm{p}}$ determined $\Omega_b h^2$'s are between $1.4\sigma$ and $2.0\sigma$ lower than the corresponding cosmological data determined values, in agreement with the Z18 findings. However, for the flat models the All 15 median statistics determined $\Omega_b h^2$'s are in very good agreement with corresponding cosmology data determined $\Omega_b h^2$'s. 

Interestingly, we find that for the nonflat cosmogonies, the All 15 median statistics (D/H)$_{\rm{p}}$-determined $\Omega_b h^2$'s are all about $2.1\sigma$ lower than what the cosmology data favor: the observed (D/H)$_{\rm{p}}$ abundance data favor flat spatial hypersurfaces over positively curved ones --- given the cosmological constraints --- at just above $2\sigma$ significance. It would be interesting to include the (D/H)$_{\rm{p}}$ measurements in a full likelihood analysis  with the other cosmological data, along the lines of \cite{Park2018b, Park2018c}; this will need to be done to more carefully weigh the consequences of our findings here. However, we can now qualitatively add the (D/H)$_{\rm{p}}$ measurements to the milder evidence from reionization \citep{Mitra2018} and the stronger evidence from the shape of the smaller-scale CMB anisotropy \citep{Ooba2018a, Ooba2017a, Ooba2017b, Ooba2018b,  Park2018a, Park2018b, Park2018c} that favor flat over closed spatial hypersurfaces, while the larger-scale CMB anisotropy shape and weak lensing measurements \citep{DESCollab2017} favor closed over flat cosmogonies \citep{Ooba2018a, Ooba2017a, Ooba2017b, Ooba2018b, Park2018a, Park2018b, Park2018c}.

\section{\label{conclusion}\textbf{CONCLUSION}}

Our median statistics analysis of the complete set of 15 (D/H)$_{\rm{p}}$ measurements compiled by Z18 results in an $\Omega_b h^2$ estimate that is very consistent with those estimated from cosmological data in spatially-flat cosmogonies, but is about $2\sigma$ lower than what cosmology data favor in closed models. A full likelihood analysis including other cosmological data will need to be performed in order to determine the proper significance of this result.

\section{\label{acknowledgements}\textbf{Acknowledgements}}

We acknowledge valuable discussions with Tia Camarillo and Aman Singal. This project was supported by funding from Kansas State University's REU program funded by the NSF grant PHY-1461251. BR was supported in part by DOE grant DE-SC0011840.

\end{document}